\documentclass[12pt,english]{article}
\usepackage[T1]{fontenc}
\usepackage[latin9]{inputenc}
\usepackage{geometry}
\usepackage{color}
\usepackage{array}
\usepackage{float}
\usepackage{graphicx}

\makeatletter

\providecommand{\tabularnewline}{\\}

\usepackage{babel}
\usepackage{times}

\usepackage{cite}

\makeatother

\usepackage{babel}
\begin{document}

\title{Semi-quantum communication: Protocols for key agreement, controlled
secure direct communication and dialogue}

\author{{\normalsize{}Chitra Shukla$^{a,}$}\thanks{email: shukla.chitra@i.mbox.nagoya-u.ac.jp }{\normalsize{},
Kishore Thapliyal$^{b,}$}\thanks{email: tkishore36@yahoo.com}{\normalsize{},
Anirban Pathak$^{b,}$}\thanks{email: anirban.pathak@jiit.ac.in}{\normalsize{}}\\
{\normalsize{}$^{a}$Graduate School of Information Science, Nagoya
University, Furo-cho 1, }\\
{\normalsize{}Chikusa-ku, Nagoya, 464-8601, Japan }\\
{\normalsize{}$^{b}$Jaypee Institute of Information Technology, A-10,
Sector-62, Noida, UP-201307, India }\\
{\normalsize{} }}
\maketitle
\begin{abstract}
Semi-quantum protocols that allow some of the users to remain classical
are proposed for a large class of problems associated with secure
communication and secure multiparty computation. Specifically, first
time semi-quantum protocols are proposed for key agreement, controlled
deterministic secure communication  and dialogue, and it is shown
that the semi-quantum protocols for controlled deterministic secure
communication and dialogue can be reduced to semi-quantum protocols
for e-commerce and private comparison (socialist millionaire problem),
respectively. Complementing with the earlier proposed semi-quantum
schemes for key distribution, secret sharing and deterministic secure
communication, set of schemes proposed here and subsequent discussions
have established that almost every secure communication and computation
tasks that can be performed using fully quantum protocols can also
be performed in semi-quantum manner. Further, it addresses a fundamental
question in context of a large number problems-\textcolor{red}{{} }how
much quantumness is (how many quantum parties are) required to perform
a specific secure communication task? Some of the proposed schemes
are completely orthogonal-state-based, and thus, fundamentally different
from the existing semi-quantum schemes that are conjugate-coding-based.
Security, efficiency and applicability of the proposed schemes have
been discussed with appropriate importance.
\end{abstract}

\section{Introduction\label{sec:Introduction}}

Since Bennett and Brassard's pioneering proposal of unconditionally
secure quantum key distribution (QKD) scheme based on conjugate coding
\cite{bb84}, various facets of secure communication have been explored
using quantum resources. On the one hand, a large number of conjugate-coding-based
(BB84-type) schemes \cite{book,ekert,b92} have been proposed for
various tasks including QKD \cite{book,ekert,b92}, quantum key agreement
(QKA) \cite{QKA_our}, quantum secure direct communication (QSDC)
\cite{ping-pong,Anindita}, deterministic secure quantum communication
(DSQC) \cite{dsqc-1,review}, quantum e-commerce \cite{online-shop},
quantum dialogue \cite{ba-an,baan_new,Man,qd,Naseri,shi-auxilary,xia,dong-w,gao-swapping,QD-EnSwap,referee2,QD-qutrit,com-QD-qut,CV-QD,probAuthQD,QSDD1,QSDD2,quantum_telephon1,Y_Sun_improve_telephone,sakshi-panigrahi-epl},
etc., on the other hand, serious attempts have been made to answer
two extremely important foundational questions- (1) Is conjugate coding
necessary for secure quantum communication? (2) How much quantumness
is needed for achieving unconditional security? Alternatively, whether
all the users involved in a secure communication scheme are required
to be quantum in the sense of their capacity to perform quantum measurement,
prepare quantum states in more than one mutually unbiased basis (MUBs)
and/or the ability to store quantum information? Efforts to answer
the first question have led to a set of orthogonal-state-based schemes
\cite{QKA_our,N09,vaidman-goldenberg,cs-thesis,QPC_Kishore}, where
security is obtained without using our inability to simultaneously
measure a quantum state using two or more MUBs. These orthogonal-state-based
schemes \cite{QKA_our,N09,vaidman-goldenberg,cs-thesis,QPC_Kishore}
have strongly established that any cryptographic task that can be
performed using a conjugate-coding-based scheme can also be performed
using an orthogonal-state-based scheme. Similarly, efforts to answer
the second question have led to a few semi-quantum schemes for secure
communication which use lesser amount of quantum resources than that
required by their fully quantum counterparts (protocols for same task
with all the participants having power to use quantum resources).
Protocols for a variety of quantum communication tasks have been proposed
under the semi-quantum regime; for example, semi-quantum key distribution
(SQKD) \cite{Talmor2007b,talmor2,Zou_less_than_4_quantum_states,QKD-limited-c-Bob,Semi-AQKD,Zou2015qinp,nearly-c-Bob,MediatedSQKD,multiuserSemiQKD,Li16_without_classical_channel,QKD_with_classical_Alice,Kraweck-phd-thesis-2015,Krawec-3-qinp,tamor3,goutampal-eavesdropping},
semi-quantum information splitting (SQIS) \cite{nie-sqis1}, semi-quantum
secret sharing (SQSS) \cite{LI-sqss,sqss-2010,attack_on_sqis}, semi-quantum
secure direct communication (SQSDC) \cite{semi-AQSDC,3-step-QSDC},
semi-quantum private comparison \cite{SQPC,QPC_Kishore}, authenticated
semi-quantum direct communication \cite{semi-AQSDC} have been proposed.
The majority of these semi-quantum schemes are two-party schemes,
but a set of multi-party schemes involving more than one classical
Bob have also been proposed \cite{multiuserSemiQKD}. In some of these
multi-party semi-quantum schemes (especially, for multiparty SQKD)
it has been assumed that there exist a completely untrusted server/center
Charlie who, is a quantum user and either all \cite{MediatedSQKD}
or some \cite{nearly-c-Bob} of the other users are classical. Further,
some serious attempts have been made for providing security proof
for semi-quantum protocols \cite{1Security-SQKD,2Security-SQKD,3Security-SQKD,security_zhang1,security_zhang2,Krwawec-2-2015-security-proof}.
However, to the best of our knowledge until now no semi-quantum protocol
has been proposed for a set of cryptographic tasks, e.g., (i) semi-quantum
key agreement (SQKA), (ii) controlled deterministic secure semi-quantum
communication (CDSSQC), (iii) semi-quantum dialogue (SQD). These tasks
are extremely important for their own merit as well as for the fact
that a scheme of CDSSQC can be easily reduced to a scheme of semi-quantum
e-commerce in analogy with Ref. \cite{e-commerce}, where it is shown
that a controlled-DSQC scheme can be used for designing a scheme for
quantum online shopping. Further, a scheme for online shopping will
be of much more practical relevance if end users (especially buyers)
do not require quantum resources and consequently can be considered
as classical users. In brief, a semi-quantum scheme for e-commerce
is expected to be of much use. It is also known that a Ba An type
scheme for QD \cite{ba-an,qd} can be reduced to a scheme of QPC \cite{qd,QPC_Kishore}
and then the same can be used to solve socialist millionaire problem
\cite{qd}; a scheme of QKA can be generalized to multiparty case
and used to provide semi-quantum schemes for sealed bid auction \cite{Our-auction};
and in a similar manner, a CDSSQC scheme can be used to yield a scheme
for semi-quantum binary quantum voting in analogy with \cite{Voting1,Voting2}.
The fact that no semi-quantum scheme exists for SQD, SQKA and CDSSQC
and their wide applicability to e-commerce, voting, private comparison
and other cryptographic tasks have motivated us to design new protocols
for SQD, SQKA and CDSSQC and to critically analyze their security
and efficiency. To do so, we have designed 2 new protocols for CDSSQC
and one protocol each for SQD and SQKA. These new protocols provide
some kind of completeness to the set of available semi-quantum schemes
and allows us to safely say that any secure communication task that
can be performed using full quantum scheme can also be performed with
a semi-quantum scheme. Such reduction of quantum resources is extremely
important as quantum resources are costly and it is not expected that
all the end users would possess quantum devices.

Before we proceed further, it would be apt to note that in the existing
semi-quantum schemes different powers have been attributed to the
classical party (lets call him as Bob for the convenience of the discussion,
but in practice we often name a classical user as Alice, too). Traditionally,
it is assumed that a classical Bob does not have a quantum memory
and he can only perform a restricted set of classical operations over
a quantum channel. Specifically, Bob can prepare new qubits only in
the classical basis (i.e., in Z basis or $\{|0\rangle,|1\rangle\}$
basis). In other words, he is not allowed to prepare $|\pm\rangle$
or other quantum states that can be viewed as superposition of $|0\rangle$
and $|1\rangle$ states. On receipt of a qubit Bob can either resend
(reflect) the qubit (independent of the basis used to prepare the
initial state) without causing any disturbance or measure it only
in the classical basis. He can also reorder the sequence of qubits
received by him by sending the qubits through different delay lines.
In fact, first ever semi-quantum scheme for key distribution was proposed
by Boyer et al., in 2007 \cite{Talmor2007b}. In this pioneering work,
the user with restricted power was referred to as classical Bob and
in that work and in most of the subsequent works  (\cite{MediatedSQKD,Semi-AQKD,semi-AQSDC,Zou_less_than_4_quantum_states}
and references therein) it was assumed that Bob has access to a segment
of the quantum channel starting from Alice's lab going back to her
lab via Bob's lab; as before, the classical party Bob can either leave
the qubit passing through the channel undisturbed or perform measurement
in the computational basis, which can be followed by fresh preparation
of qubit in the computational basis. This was followed by a semi-quantum
scheme of key distribution \cite{talmor2}, where the classical party
can either choose not to disturb the qubit or to measure it in the
computational basis, and instead of preparing the fresh qubits in
the computational basis he may reorder the undisturbed qubits to ensure
unconditional security. Later, these schemes were modified to a set
of SQKD schemes with less than four quantum states, where Alice requires
quantum registers in some of the protocols \cite{Zou_less_than_4_quantum_states},
but does not require it in all the protocols. In what follows, we
attribute the same power to Bob in the schemes proposed in this paper.
In the schemes proposed below, senders are always classical and they
are referred to as classical Bob/Alice. This is in consistency with
the nomenclature used in most of the recent works  (\cite{Talmor2007b,talmor2,MediatedSQKD,Semi-AQKD,semi-AQSDC,Zou_less_than_4_quantum_states}
and references therein). However, in the literature several other
restrictions have been put on classical Bob. For example, in Ref.
\cite{QKD-limited-c-Bob} a SQKD scheme was proposed where Bob was
not allowed to perform any measurement and he was referred to as ``limited
classical Bob''. It was argued that a limited classical Bob can circumvent
some attacks related to the measurement device \cite{Implementation_attack1,Implementation_attack2,Implementation_attack3}.
However, the scheme proposed in \cite{QKD-limited-c-Bob} was not
measurement device independent. Similarly, in \cite{nearly-c-Bob}
a server was delegated the task of performing measurement and application
of one of the two Pauli operations, while the classical users role
was restricted in randomly sending the received qubits to the server
for random application of operator or measurement. Such a classical
user was referred to as ``nearly classical Bob''. 

Remaining part of the paper is organized as follows. In Sec. \ref{sec:Two-party-semi-quantum-keyA},
a protocol for SQKA among a quantum and a classical user is proposed.
Two CDSSQC schemes with a classical sender and a receiver and controller
both possessing quantum powers are proposed in Sec. \ref{sec:Controlled-direct-secureSQC}.
In Sec. \ref{sec:Semi-quantum-dialogue}, a SQD scheme between a classical
and a quantum parties are designed. The security of the proposed schemes
against various possible attacks are discussed in the respective sections.
The qubit efficiency of the proposed schemes has been calculated in
Sec. \ref{sec:Efficiency-analysis} before concluding the work in
Sec. \ref{sec:Conclusion}.

\section{Protocol for semi-quantum key agreement \label{sec:Two-party-semi-quantum-keyA}}

In analogy of the weaker notion of quantum key agreement, i.e., both
the parties take part in preparation of the final shared key, most
of the SQKD protocols may be categorized as SQKA schemes. Here, we
rather focus on the stronger notion of the key agreement, which corresponds
to the schemes where each party contribute equally to the final shared
key, and none of the parties can manipulate or know the final key
prior to the remaining parties (for detail see \cite{QKA_our} and
references therein). We will also show that the proposed SQKA scheme
can be reduced to a scheme for SQKD, first of its own kind, in which
a sender can send a quantum key to the receiver in an unconditionally
secure and deterministic manner.

\begin{figure}[H]
\begin{centering}
\includegraphics[scale=0.7]{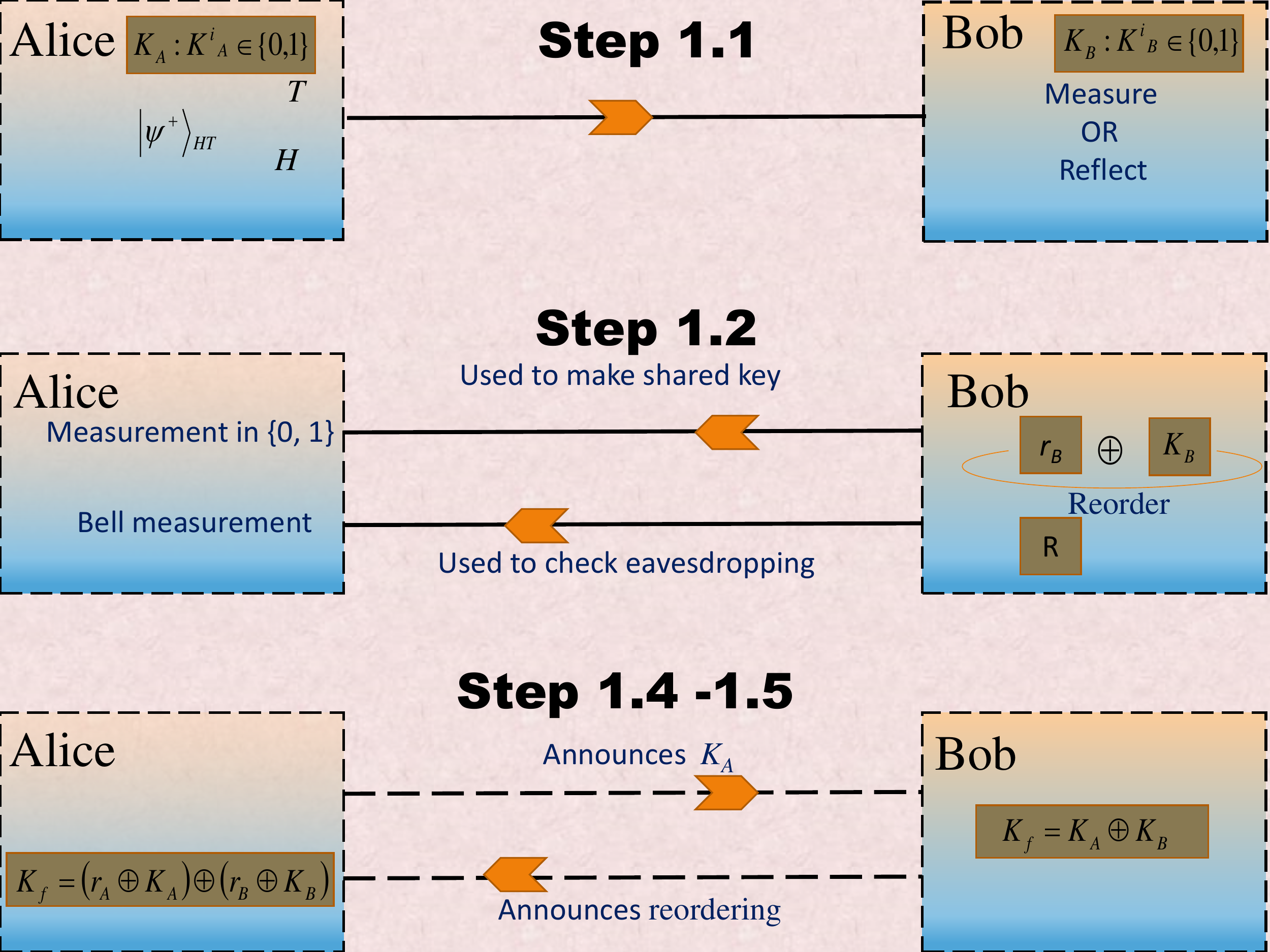}
\par\end{centering}

\caption{\label{fig:Schematic}(Color online) Schematic diagram illustrating
the working of protocol for SQKA proposed here. The protocol is illustrated
through three steps and is explained through Step1.1-Step1.5.}
\end{figure}

In this section, we propose a two-party semi-quantum key agreement
protocol, where Alice has all quantum powers, but Bob is classical,
i.e., he is restricted to perform the following operations (1) measure
and prepare the qubits only in the computational basis $\{|0\rangle,|1\rangle\}$
(also called classical basis), and (2) simply reflects the qubits
without disturbance. The working of the proposed protocol is illustrated
through a schematic diagram shown in Fig. \ref{fig:Schematic}. 

The following are the main steps in the protocol.

\section*{Protocol~1\label{sec:Protocol1}}
\begin{description}
\item [{Step1.1:}] \textbf{Preparation of the quantum channel: }Alice prepares
$n+m=N$ number of Bell states, i.e., $|\psi^{+}\rangle^{\otimes N}$,
where $|\psi^{+}\rangle=\frac{|00\rangle+|11\rangle}{\sqrt{2}}$.
Out of the total $N$ Bell states, $n$ will be used for key generation
while the remaining $m$ will be used as decoy qubits for eavesdropping
checking. Subsequently, she prepares two ordered sequences of all
the first qubits and all the second qubits from the initial Bell states.
She keeps the first sequence with herself as home qubits (H) and sends
the second sequence to Bob as travel qubits (T) as shown by arrow
in Step1.1 of Fig. \ref{fig:Schematic}. Using a quantum random number
generator (QRNG), Alice also prepares her raw key of $n$ bits $K_{A}=\{K_{A}^{1},K_{A}^{2}\cdots K_{A}^{i}\cdots K_{A}^{n}\}$,
where $K_{A}^{i}$ is the $i^{th}$ bit of $K_{A}$, and $K_{A}^{i}\in\{0,1\}$
as shown in Step1.1 of Fig. \ref{fig:Schematic}.\textcolor{red}{{} }
\item [{Step1.2:}] \textbf{Bob's encoding:} Bob also prepares his raw key
$K_{B}=\{K_{B}^{1},K_{B}^{2},\cdots,K_{B}^{i},\cdots,K_{B}^{n}\}$
of $n$ bits by using a RNG\footnote{Bob being classical, we refer to his random number generator as a
traditional pseudo random number generator, instead of a true random
number generator which is required to be quantum. This aspect of random
key generation in Bob's side is not explicitly discussed in the existing
works on semi-quantum protocols. However, it's not a serious issue
as on one hand, extremely good quality pseudo random numbers can be
generated classically (thus, use of a classical RNG will be sufficient
for the present purpose); on the other hand, QRNG are now commercially
available and is not considered as a costly quantum resource, so if
one just allows an otherwise classical user to have an QRNG, the modified
scheme could still be considered as a semi-quantum scheme as Bob would
still lack the power of performing quantum measurement and or storing
quantum information. In fact, measurement in classical basis would
be sufficient for generation of true random number if Bob can create
a $|\pm\rangle$ state using a beam splitter.}, (which is independent of Alice's QRNG), where $K_{B}^{i}$ is the
$i^{th}$ bit of the $K_{B}$ with $K_{B}^{i}\in\{0,1\}$. After receiving
all the qubits from Alice, Bob randomly chooses one of the two operations,
either to measure or reflect as shown in Step1.1 of Fig. \ref{fig:Schematic}.
Specifically, he measures $n$ qubits (chosen randomly) in the computational
basis, while reflects the remaining $m$ qubits to be used later for
eavesdropping checking. \\
He forms a string of his measurement outcomes as $r_{B}=\{r_{B}^{1},r_{B}^{2},\cdots,r_{B}^{i},\cdots,r_{B}^{n}\}$,
where $r_{B}^{i}$ is the measurement outcome of $i^{th}$ qubit chosen
to be measured by Bob in the computational basis, and therefore, $r_{B}^{i}\in\{0,1\}$.
Then to encode his raw key $K_{B}$, he performs a bit wise XOR operation,
i.e., $r_{B}\oplus K_{B}$ and prepares corresponding qubits in the
computational basis. Finally, he inserts the encoded $n$ qubits back
into the string of reflected $m$ qubits and sends the resultant sequence
$r^{b}$ back to Alice only after applying a permutation operator
$\Pi_{N}$ as shown by the first arrow in Step1.2 of Fig. \ref{fig:Schematic}.
These qubits would further be used to make the final shared key $K_{f}.$
\item [{Step1.3:}] \textbf{Announcements and eavesdropping checking:} After
receiving an authenticated acknowledgment of the receipt of all the
qubits from Alice, Bob announces the permutation operator $\Pi_{m}$
corresponding to the qubits reflected by him. Though, this would reveal
which qubits have been measured and which qubits have been reflected
by Bob but Eve or Alice cannot gain any advantage due to lack of information
regarding the permutation operator $\Pi_{n}$. Further, to detect
eavesdropping, Alice firstly measures the reflected qubits in Bell
basis by combining them with respective partner home qubits. If she
finds the measurement result as $|\psi^{+}\rangle$ state then they
confirm that no eavesdropping has happened, because the initial state
was prepared in $|\psi^{+}\rangle$ state, and they move on to the
next step, otherwise they discard the protocol and start from the
beginning.
\item [{Step1.4:}] \textbf{Extraction of the final shared key $K_{f}$
by Bob:} After ensuring that there is no eavesdropping, Alice announces
her secret key $K_{A}$ publicly as shown by the first arrow in Step1.4
of Fig. \ref{fig:Schematic}, and Bob uses that information to prepare
his final shared key $K_{f}=K_{A}\oplus K_{B}$ as he knows $K_{B}$.
Subsequently, he also reveals the the permutation operator $\Pi_{n}$.
\\
Here, it is important to note that Alice announces her secret key
$K_{A}$ only after the receipt of all the encoded qubits from Bob.
So, now Bob can not make any further changes as per his wish.
\item [{Step1.5:}] \textbf{Extraction of the final shared key $K_{f}$
by Alice:} Once Alice has reordered Bob's encoded qubits she measures
both the home (H) and travel (T) qubits in the computational basis.
She obtains a string of $n$ bits $r_{A}$ corresponding to measurement
outcomes of home qubits, and she can use the measurement outcomes
of the travel qubits to know Bob's encoded raw key as the initial
Bell states are publicly known. For the specific choice of initial
Bell state here, i.e., $|\psi^{+}\rangle$, the relation $r_{A}=r_{B}$
holds. Therefore, the final shared key would be 
\end{description}
\begin{equation}
\begin{array}{lcl}
K_{f} & = & K_{A}\oplus K_{B}=(r_{A}\oplus K_{A})\oplus(r_{B}\oplus K_{B}).\end{array}\label{eq:Final-Key}
\end{equation}
Hence, the final shared key $K_{f}$ is shared between Alice and Bob.

\begin{table}[H]
\begin{centering}
\begin{tabular}{|>{\centering}p{3.5cm}|>{\centering}p{3.5cm}|>{\centering}p{3cm}|>{\centering}p{3.5cm}|}
\hline 
Alice's measurement result $r_{A}$ on home (H) qubits & Bob's measurement result $r_{B}$ & Bob's secret key $K_{B}$ & Alice's measurement result on travel (T) qubits\tabularnewline
\hline 
$|0\rangle$ & $|0\rangle$ & $|0\rangle$ & $|0\rangle$\tabularnewline
\hline 
 &  & $|1\rangle$ & $|1\rangle$\tabularnewline
\hline 
$|1\rangle$ & $|1\rangle$ & $|0\rangle$ & $|1\rangle$\tabularnewline
\hline 
 &  & $|1\rangle$ & $|0\rangle$\tabularnewline
\hline 
\end{tabular}
\par\end{centering}

\caption{\label{SQKA} All the possibilities during Alice's extraction of Bob's
secret key.}
\end{table}

In Eq. (\ref{eq:Final-Key}), Eve may know Alice's secret key $K_{A}$
as this was announced through a classical channel. She is also aware
of $r_{A}=r_{B}$ due to public knowledge of the initial choice of
Bell state. However, it does not affect the secrecy of the final shared
key $K_{f}$ which is prepared as $K_{A}\oplus K_{B}$, because Eve
does not know anything about Bob's secret key $K_{B}$ and the value
of $r_{A}$ (or $r_{B}$).

Further, it should be noted here the computational basis measurement
is not the only choice by Alice, rather she can extract Bob's encoding
by performing Bell measurement. Here, we will skip that discussion
as the same has been discussed in the following section for semi-quantum
dialogue protocol.

If we assume that Alice is not intended to send her raw key (i.e.,
she does not announce her raw key) in the proposed SQKA protocol,
indeed following it faithfully, then it will reduce to a deterministic
SQKD protocol. Specifically, in analogy of ping-pong protocol \cite{ping-pong}
to perform a quantum direct communication task, which was also shown
to share a quantum key in a deterministic manner.

\subsection{Possible attack strategies and security \label{sub:Possible-attack-strategiesQKA}}
\begin{enumerate}
\item \textbf{Eve's attack: }As\textbf{ }mentioned beforehand Eve's ignorance
regarding the final shared key solely depends on the fact whether
Alice receives Bob's raw key in a secure manner. In other words, although
Eve is aware of the initial state and Alice's raw key, she still requires
Bob's key to obtain the final shared key. In what follows, we will
discuss some attacks, she may attempt to extract this information.
\\
The easiest technique Eve may incorporate is a \emph{CNOT attack}
(as described and attempted in Refs. \cite{Talmor2007b,talmor2,talmor2007}).
To be specific, she may prepare enough ancilla qubits (initially as
$|0\rangle$) to perform\textcolor{blue}{{} }a CNOT with each travel
qubit at control and an ancilla as a target while Alice to Bob communication.
This way the compound state of Alice's, Bob's and Eve's qubits, prior
to Bob's measurement, becomes $|\psi_{{\rm ABE}}\rangle=\frac{|000\rangle+|111\rangle}{\sqrt{2}}$.
As Bob returns some of the qubits performing single qubit measurement
in the computational basis on his qubit (B). The reduced state of
Alice's and Eve's qubits may be written as $|\rho_{{\rm AE}}\rangle=\frac{|00\rangle\langle00|+|11\rangle\langle11|}{2}$
corresponding to Bob's measurement, while the three qubit state remains
unchanged for reflected qubits. Suppose Bob prepares a fresh qubit
$|\xi_{{\rm B}}\rangle=|r_{B}\oplus K_{B}\rangle$ and returns the
string of encoded qubits (in other words, measured qubits) and reflected
qubits to Alice (without applying a permutation operator). Subsequently,
Eve again performs a CNOT operation while Bob to Alice communication
with control on travel qubits and target on the ancilla qubits. It
is straightforward to check that in case of reflected qubits the state
reduces to $|\psi_{{\rm ABE}}\rangle=\frac{|00\rangle+|11\rangle}{\sqrt{2}}\otimes|0\rangle$.
Whereas, for encoded qubits it may be written as $|\rho_{{\rm ABE}}^{\prime}\rangle=\frac{1}{2}\left(|0\rangle\langle0|\otimes|K_{B}\rangle\langle K_{B}|+|1\rangle\langle1|\otimes|K_{B}\oplus1\rangle\langle K_{B}\oplus1|\right)\otimes|K_{B}\rangle\langle K_{B}|$.
From which it may be observed that Eve will always obtain Bob's secret
key. However, this problem is circumvented using a permutation operator
(in Step1.2) by Bob on the string of encoded and reflected qubits.\\
As Eve's CNOT attack strategy is foiled by the use of a permutation
operator she may attempt other attack strategies. Suppose she performs
an \emph{intercept and resend attack}. Specifically, in this attack,
she can prepare an equal number of Bell states as Alice and send all
the second qubits to Bob, keeping the Alice's original sequence with
herself. Bob follows the protocol and encodes $n$ qubits randomly
and sends them to Alice, which is again intercepted by Eve. Subsequently,
Eve performs the Bell measurement on all the Bell pairs (which she
had initially prepared), and she may come to know which $n$ qubits
were measured by Bob. Quantitatively, she can get this knowledge $75\%$
of the time as in the Bell measurement outcomes anything other than
the original state would result in a measurement performed by Bob.
Depending upon the Bell measurement outcomes Bob's encoding can also
be revealed as $|\psi^{-}\rangle$ and $|\phi^{\pm}\rangle$ will
correspond to Bob's 0 and 1 in the computational basis, respectively
(see Section \ref{sec:Semi-quantum-dialogue} and Table \ref{tab:QD-table}
for more detail). Subsequently, she performs a measurement in the
computational basis on the qubits sent by Alice corresponding to each
qubit Bob has measured. Finally, she sends the new string of qubits
(comprising of freshly prepared and Alice's original qubits) to Alice
which will never fail in eavesdropping checking and Alice will announce
her key and Eve can get at least 75\% of the shared key.\\
It is important to note here that 25\% of the key of Alice and Bob
will also not match in this case. This may be attributed to the disturbance
caused due to eavesdropping, which left that signature and is a characteristic
nature of quantum communication. This fact can be explored to achieve
the security from this kind of an attack. Specifically, Alice and
Bob may choose to perform a verification strategy of a small part
of the shared key to check this kind of an attempt. \\
As we have already incorporated a permutation of particles scheme
(performed by Bob) for security against CNOT attack, it becomes relevant
to see the feasibility of this attack in this case as well. Bob discloses
the permutation operator for decoy qubits only after an authenticated
acknowledgment by Alice. Therefore, Eve fails to obtain the encoded
bit value prior to this disclosure of Bob as it is, although it's
encoded in the computational basis, but she does not know the partner
Bell pair due to randomization. She will require this Bell pair to
decode the information as the measurement outcome of the partner Bell
particle acts as a key for decoding the Bob's information. Further,
Bob announces the correct order of particles only when less than a
threshold of errors are found during eavesdropping checking. \\
Indeed, most of the attacks by an eavesdropper can be circumvented
if the classical Bob is given power to permute the string of qubits
with him, i.e., Bob can secure his raw key (information) in the proposed
SQKA scheme by permuting the particles before sending to Alice. \\
There are some other attacks (see \cite{AQD,QC} for details) which
do not affect the security of the proposed protocol, like \emph{disturbance
attack}, \emph{denial of service attack}, and \emph{impersonation
attack}\textbf{ }(as it becomes void after incorporating an authentication
protocol).
\item \textbf{Alice's attack: }In the eavesdropping checking, at the end
of the round trip of Bell pairs, Bob announces the positions of the
reflected qubits in each of the Bell pairs and the remaining string
(i.e., encoded string after measurement) is in the computational basis
and Alice can know Bob's encoding before she announces her own. In
other words, she can control the final key completely as she can announce
her raw key accordingly. However, this is not desired in a genuine
key agreement scheme.\\
This possible attack by Alice is circumvented by the use of permutation
operator discussed in the last attack. As Bob reveals the permutation
he had applied on the freshly prepared qubits (on which his raw key
is encoded) only after Alice announces her raw key, she can not extract
his raw key due to lack of knowledge of pair particles corresponding
to each initially prepared Bell state. Hence, only after Alice's announcement
of her raw key she comes to know Bob's raw key with his cooperation.
\\
To avoid this attack, we may also decide that both Alice and Bob share
the hash values of their raw keys during their communication due to
which if she wishes to change her raw key later, then the protocol
is aborted as the hash value for her modified raw key will not match
with original raw key.
\item \textbf{Bob's attack: }As mentioned in the Alice's attack that Bob
announces the permutation operator only after receiving her raw key.
One should notice here that the permuted string Bob has sent and corresponding
Alice's string are in computational basis. Further, Bob knows each
bit value in Alice's string as those are nothing but Bob's corresponding
measurement outcomes in Step 1.2. Once Bob knows Alice's raw key,
he may control the final key entirely by disclosing a new permutation
operator that suits his choice of shared key.\\
Therefore, it becomes important to incorporate the hash function as
if Bob has already shared the hash value of his key he cannot change
his raw key during announcement of permutation operator.
\end{enumerate}

\section{Controlled direct secure semi-quantum communication \label{sec:Controlled-direct-secureSQC}}

If we observe the SQKA protocol proposed here, it can be stated that
Bob sends his raw key by a DSSQC scheme and Alice announces her raw
key. The security of the final key depends on the security of the
raw key of Bob. Hence, a semi-quantum counterpart of direct communication
scheme can be designed. However, avoiding the designing of various
schemes for the same task, we rather propose a controlled version
of direct communication scheme and discuss the feasibility of realizing
this scheme, which would directly imply the possibility of direct
communication scheme. Here, we will propose two controlled direct
secure semi-quantum communication protocols. Note that in the proposed
CDSSQC schemes only Alice is considered as a\textcolor{red}{{} }classical
party, while Bob and Charlie possess quantum powers.

\subsection{Protocol 2: Controlled direct secure semi-quantum communication }

The working of this scheme is as follows.
\begin{description}
\item [{Step2.1:}] \textbf{Preparation of shared quantum channel: }Charlie
prepares $n+m=N$ copies of a three qubit entangled state\textbf{
}
\begin{equation}
|\psi\rangle_{GHZ-{\rm like}}=\frac{|\psi_{1}\rangle|a\rangle+|\psi_{2}\rangle|b\rangle}{\sqrt{2}},\label{eq:CDSQC-channel}
\end{equation}
where $|\psi_{i}\rangle\in\left\{ |\psi^{+}\rangle,|\psi^{-}\rangle,|\phi^{+}\rangle,|\phi^{-}\rangle:|\psi^{\pm}\rangle=\frac{|00\rangle\pm|11\rangle}{\sqrt{2}},\,|\phi^{\pm}\rangle=\frac{|01\rangle\pm|10\rangle}{\sqrt{2}},\right\} $
and $\langle a|b\rangle=\delta_{a,b}$. The classical user Alice will
encode her $n$-bit message on the $n$ copies, while the remaining
$m$ copies will be used as decoy qubits to check an eavesdropping
attempt. Subsequently, Charlie prepares three sequences of all the
first, second and third qubits of the entangled states. Finally, he
sends the first and second sequences to Alice and Bob, respectively.
\textcolor{red}{}\\
They can check the correlations in a few of the shared quantum states
to avoid an eavesdropping attempt using intercept and resend attack,
i.e., Charlie measures his qubits in $\left\{ |a\rangle,|b\rangle\right\} $
basis, while Alice and Bob in computational basis. However, such an
eavesdropping test would fail to provide security against measurement
and resend attack. Security against such an attack is discussed later.\\
In addition, Charlie and Bob both being capable of performing quantum
operations may perform BB84 subroutine (cf. \cite{Voting1,decoy,AQD}
and references therein) to ensure a secure transmission of the qubits
belonging to quantum channel. This would provide additional security
against intercept-resend attacks\textcolor{red}{{} }on  Charlie-Bob
quantum channel.
\item [{Step2.2:}] \textbf{Alice's encoding:} Alice has a $n$ bit message
$M=\{M_{A}^{1},M_{A}^{2}\cdots M_{A}^{i}\cdots M_{A}^{n}\}$. To encode
this message Alice measures $n$ qubits (chosen randomly) in computational
basis to obtain measurement outcomes $r_{A}=\{r_{A}^{1},r_{A}^{2}\cdots r_{A}^{i}\cdots r_{A}^{n}\}$,
and prepares a new string of qubits in $\left\{ |0\rangle,|1\rangle\right\} $
basis corresponding to bit values $M_{A}^{i}\oplus r_{A}^{i}$. Finally,
she reinserts all these qubits back into the original sequence and
sends it to Bob only after permuting the string. It is important that
she leaves enough qubits undisturbed so that those qubits may be employed
as decoy qubits.
\item [{Step2.3:}] \textbf{Announcements and eavesdropping checking:} After
receiving an authenticated acknowledgement of the receipt of all the
qubits from Bob, Alice announces which qubits have been encoded and
which qubits have been left as decoy qubits. She also discloses the
permutation operator applied only on the decoy qubits. Further, to
detect eavesdropping, Bob firstly measures the pair of decoy qubits
from Alice's and Bob's sequences in the Bell basis and with the help
of Charlie's corresponding measurement outcome (which reveals the
initial Bell state Alice and Bob were sharing) he can calculate the
errors. If sufficiently low errors are found they proceed to the next
step, otherwise start afresh.
\item [{Step2.4:}] \textbf{Decoding the message:} To decode the message,
Bob can perform a measurement in the computational basis on all the
remaining qubits from both the sequences received from Charlie and
Alice. Subsequently, Alice also discloses her permutation on the message
encoded (or freshly prepared) qubits in her string. However, Bob cannot
 decode Alice's secret message yet, as he remains unaware of the Bell
state he was sharing with Alice until Charlie announces his measurement
outcome. 
\item [{Step2.5:}] \textbf{Charlie's announcement:} Finally, Charlie announces
his measurement outcome in $\left\{ |a\rangle,|b\rangle\right\} $
basis using which Bob can decode Alice's message.
\end{description}

\subsection{Protocol 3: Controlled direct secure semi-quantum communication based
on cryptographic switch}

This controlled communication scheme is based on quantum cryptographic
switch scheme proposed in the past \cite{switch} and has been shown
to be useful in almost all the controlled communication schemes \cite{Voting1,crypt-switch,cdsqc,referee1}.
\begin{description}
\item [{Step3.1:}] \textbf{Preparation of the shared quantum channel: }Charlie
prepares $n+m=N$ copies of one of the Bell states, out of which $n$
Bell pairs will be used for sending messages and the rest as decoy
qubits. Subsequently, Charlie prepares two sequences of all the first
and second qubits of the entangled state. He also performs a permutation
operator on the second sequence. Finally, he sends the first and second
sequences to Alice and Bob, respectively.\\
Both Alice and Bob may check the correlations in a few of the shared
Bell states to avoid an eavesdropping attempt as was done in Step2.1.
Similarly, Charlie and Bob both being capable of performing quantum
operations may also perform BB84 subroutine (cf. \cite{Voting1,decoy,AQD}
and references therein).
\item [{Step3.2:}] Same as Step2.2 of Protocol 2.
\item [{Step3.3:}] \textbf{Announcements and eavesdropping checking:} After
receiving an authenticated acknowledgment of the receipt of all the
qubits from Bob, Alice announces which qubits have been encoded and
which qubits have been left as decoy qubits. She also discloses the
permutation operator corresponding to the decoy qubits only. Then
Charlie announces the correct positions of the partner pairs of decoy
Bell states in the Bob's sequence. To detect eavesdropping, Bob measures
the pairs of decoy qubits from Alice's and Bob's sequences in the
Bell basis to calculate the errors. If sufficiently low errors are
found they proceed to the next step, otherwise start afresh.
\item [{Step3.4:}] \textbf{Decoding the message:} To decode the message
Bob can perform a measurement in the computational basis on all the
remaining qubits from both the sequences received from Charlie and
Alice. Meanwhile, Alice discloses her permutation operator enabling
Bob to decode her message. However, he cannot decode Alice's secret
message yet as he is unaware of the permutation operator Charlie has
applied. 
\item [{Step3.5:}] \textbf{Charlie's announcement:} Finally, Charlie sends
the information regarding the permutation operator to Bob, using which
Bob can decode Alice's message.
\end{description}
It is important to note that two of the three parties involved in
the CDSSQC protocols are considered quantum here. The possibilities
of minimizing the number of parties required to have quantum resources
will be investigated in the near future.

In the recent past, it has been established that the controlled counterparts
of secure direct communication schemes \cite{crypt-switch,cdsqc,referee1}
can provide solutions of a handful of real life problems. For instance,
schemes of quantum voting \cite{Voting1,Voting2} and e-commerce \cite{e-commerce}
are obtained by modifying the schemes for controlled DSQC. Here, we
present the first semi-quantum e-commerce scheme, in which Alice (buyer)
is classical, and Bob (merchant) and Charlie (online store) possess
quantum resources. Both Alice and Bob are registered users of the
online store Charlie. When Alice wishes to buy an item from Bob she
sends a request to Charlie, who prepares a tripartite state (as in
Eq. \ref{eq:CDSQC-channel} of Protocol 2) to be shared with Alice
and Bob. Alice encodes the information regarding her merchandise to
Bob and encodes it as described in Step2.2. The merchant can decode
Alice's order in Step2.5 but will deliver the order only after receiving
an acknowledgment from Charlie in Step2.5. Here, it is important to
note that in some of the recent schemes, Charlie can obtain information
about Alice's order and/or change it, which is not desired in a genuine
e-commerce scheme \cite{online-shop}. The semi-quantum e-commerce
scheme modified from the proposed CDSSQC scheme is free from such
an attack as Alice applies a permutation operator and discloses her
permutation operator only after a successful transmission of all the
travel particles. In a similar manner, another quantum e-commerce
scheme may be obtained using CDSSQC scheme presented as Protocol 3,
where the online store prepares only Bell states.

\subsection{Possible attack strategies and security \label{sub:Possible-attack-strategiesCDSQC}}

Most of the attacks on the proposed CDSQC schemes may be circumvented
in the same manner as is done in the SQKA scheme in Section \ref{sub:Possible-attack-strategiesQKA}.
Here, we only mention additional attack strategies that may be adopted
by Eve.

As discussed while security of SQKA scheme, the easiest technique
for Eve would be a CNOT attack\textcolor{magenta}{.} Specifically,
she may entangle her ancilla qubits with the travel qubits from Charlie
to Alice and later disentangle them during Alice to Bob communication.
She succeeds in leaving no traces using this attack and getting Alice's
all the information (see Section \ref{sub:Possible-attack-strategiesQKA}
for detail). However, Alice may circumvent this attack just by applying
a permutation operator on all the qubits before sending them to Bob. 

Eve may choose to perform an \emph{intercept and resend attack}. Specifically,
she can prepare an equal number of single qubits in the computational
basis as Charlie has sent to Alice and send all these qubits to Alice,
keeping the Charlie's original sequence with herself. When Alice,
Bob, and Charlie check the correlations in Step 1, they will detect
uncorrelated string with Alice corresponding to this attack.

Eve can measure the intercepted qubits in the computational basis
and prepares corresponding single qubits to resend to Alice. In this
case, Eve will not be detected during correlation checking. However,
Alice transmits the encoded qubits after permutation to Bob due to
which Eve fails to decode her message, and consequently, Eve will
be detected at Bob's port during eavesdropping checking. 

As mentioned beforehand both these attacks and the set of remaining
attacks may be circumvented due to permutation operator applied by
Alice.

\section{Semi-quantum dialogue\label{sec:Semi-quantum-dialogue}}

In this section, we propose a two-party protocol for SQD, where Alice
has all quantum powers and Bob is classical. The following are the
main steps of the protocol.
\begin{description}
\item [{Step4.1:}] \textbf{Alice's preparation of quantum channel and encoding
on it: }Alice prepares $n+m=N$ number of initial Bell states, i.e.,
$|\psi^{+}\rangle^{\otimes N}$, where $|\psi^{+}\rangle=\frac{|00\rangle+|11\rangle}{\sqrt{2}}$.
She prepares two ordered sequences of all the first qubits as home
(H) qubits and all the second qubits as travel (T) qubits from the
initial Bell states. She keeps the home (H) qubits with herself and
sends the string of travel qubits to classical Bob. The initial Bell
states and encoding schemes are publicly known and let's say that
$U_{A}$ and $U_{B}$ are the measurement operations of Alice and
Bob, respectively. 
\item [{Step4.2:}] \textbf{Bob's eavesdropping checking:} Bob informs Alice
about the reception of the travel sequence by a classical channel.
Here, we can perform an eavesdroping checking strategy as discussed
above in Step 1.1 of CDSSQC schemes in Section \ref{sec:Controlled-direct-secureSQC}
that the joint measurement in the computational basis by Alice and
Bob on the Bell pairs should be correlated. If they find error rate
more than a threshold value, then they abort the protocol and starts
from the beginning. 
\item [{Step4.3:}] \textbf{Bob's encoding:} Bob measures $n$ qubits (chosen
randomly) in the computational basis and records all the measurement
outcomes in a string of bits $r_{B}$. Then he prepares a string of
his message in binary as $M_{B}$. Finally, he prepares fresh qubits
in the computational basis for each bit value $r_{B}\oplus M_{B}$
and reinserts them in the original sequence. Then he sends the encoded
and decoy qubits back to Alice after performing a permutation on them.
Here, it is important to note that the encoding operation used by
Bob can be thought equivalent to Pauli operations $\left\{ I,X\right\} $
but they are performed classically by Bob, i.e., remaining within
his classical domain.
\item [{Step4.4:}] \textbf{Alice's eavesdropping checking:} After receiving
the authenticated acknowledgement of the receipt of all the qubits
from Alice, Bob announces the positions of decoy qubits along with
the corresponding permutation operator. Alice then measures all the
decoy qubits in the Bell basis and any measurement outcome other than
that of the initially prepared Bell state would correspond to an eavesdropping
attempt. 
\item [{Step4.5:}] \textbf{Alice's encoding and measurement: }For sufficiently
low errors in eavesdropping checking Bob also discloses the permutation
operator applied on the freshly prepared message qubits and Alice
proceeds to encoding her secret on the qubits received from Bob by
applying Pauli operations $\left\{ I,X\right\} $. Finally, she measures
the partner pairs in Bell basis and announces her measurement outcomes.
From the measurement outcomes, both Alice and Bob can extract Bob's
and Alice's message, respectively. 
\end{description}
Here, it should be noted that the Bell measurement performed in Step4.5
is not necessary, a two qubit measurement performed in the computational
basis will also work in the above mentioned case. 

\begin{table}
\begin{centering}
\begin{tabular}{|c|c|c|}
\hline 
Bob's message & Alice's message & Final Bell state measurement outcome\tabularnewline
\hline 
$|0\rangle$ & $|0\rangle$ & $|\psi^{\pm}\rangle$\tabularnewline
\hline 
$|0\rangle$ & $|1\rangle$ & $|\phi^{\pm}\rangle$\tabularnewline
\hline 
$|1\rangle$ & $|0\rangle$ & $|\phi^{\pm}\rangle$\tabularnewline
\hline 
$|1\rangle$ & $|1\rangle$ & $|\psi^{\pm}\rangle$\tabularnewline
\hline 
\end{tabular}
\par\end{centering}

\caption{\label{tab:QD-table}All possible encoding of Alice and Bob with corresponding
Bell measurement outcomes. Here, $|\psi^{+}\rangle$ is assumed to
be the initial state.}
\end{table}

In the recent past, it has been established that a scheme for quantum
dialogue can be modified to provide a solution for the quantum private
comparison, which can be viewed as a special form of socialist millionaire
problem \cite{qd}. In the semi-quantum private comparison (SQPC)
task \cite{QPC_Kishore}, two classical users wish to compare their
secrets of $n$-bits with the help of an untrusted third party possessing
quantum resources. Before performing the SQPC scheme, the untrusted
third party (Alice here) prepares a large number of copies of the
Bell states using which both ${\rm Bob}_{1}$ and ${\rm Bob}_{2}$
prepare two shared unconditionally secure symmetric strings in analogy
with the schemes described in \cite{MediatedSQKD}. They use one symmetric
string as a semi-quantum key, while the other to decide the positions
of the qubits they will choose to reflect or to measure. Specifically,
both the classical users decide to measure the $i$th qubit received
during the SQPC scheme if $i$th bit in the shared string is 1, otherwise
they reflect the qubit. Using this approach both classical users prepare
fresh qubit using the encoding operation defined in Step4.3, with
the only difference that this time the transmitted information is
encrypted by the shared key. Once Alice receives all the qubits and
measures all of them in the Bell basis. Both the classical users disclose
the string they had originally shared, using which Alice announces
the measurement outcomes corresponding to the reflected qubits. ${\rm Bob}_{i}$s
can subsequently compute the error rate from the measurement outcomes
and if the error rate is below the threshold, Alice publicly announces
1-bit secret whether both the users had the same amount of assets
or not (see \cite{QPC_Kishore} for detail). Thus, we establish that
a slight modification of Protocol 4 may lead to a new scheme for SQPC.
This is interesting as to the best of our knowledge until now there
exist only two proposals for SQPC \cite{QPC_Kishore,SQPC}.\textcolor{blue}{{} }

\subsection{Possible attack strategies and security \label{sub:Possible-attack-strategiesSQD}}

Most of the attacks on the proposed SQKA and CDSSQC schemes will also
be valid on the SQD scheme. Further, as mentioned beforehand most
of these attacks will be circumvented due to permutation operator
Bob has applied. Here, it is worth mentioning that permutation operator
is not a unique way to circumvent these attacks, a prior shared key
will also ensure the security of the protocol. A similar strategy
of using a prior shared key has been observed in a few protocols in
the past \cite{QD-EnSwap}. We would like to emphasize here that employing
a key for security is beyond the domain of direct communication. Therefore,
we have preferred permutation of particles over a key in all the proposed
schemes.

Further, it is shown in the past by some of the present authors \cite{AQD,QC}
that if the information regarding the initial state is not a public
knowledge and sent using a QSDC/DSQC protocol, then an inherent possible
attack in QD schemes, i.e., information leakage attack, can be circumvented.

\section{Efficiency analysis \label{sec:Efficiency-analysis}}

Performance of a quantum communication protocol can be characterized
using qubit efficiency \cite{eff}, $\eta=\frac{c}{q+b}$, which is
the ratio of $c$-bits of message transmitted using $q$ number of
qubits and $b$-bits of classical communication. Note that the qubits
involved in eavesdropping checking as decoy qubits are counted while
calculating qubit efficiency, but the classical communication associated
with it is not considered.

Before computing the qubit efficiency of the four protocols proposed
here, we would like to note that in all the protocols the classical
senders are sending $n$ bits of secret, while in Protocol 4 (protocol
for SQD, where both classical and quantum users transmit information),
the quantum user Alice was also able to send the same amount of information
to classical Bob. In all these cases, the classical sender encodes
$n$ bits secret information using $n$-qubits. However, to ensure
the secure transmission of those $n$-qubits, another $3n$-qubits
are utilized (i.e., $m=3n$). This is so because to ensure a secure
communication of $n$ qubits, an equal number of decoy qubits are
required to be inserted randomly \cite{nielsen}. The error rate calculated
on the decoy qubits decides, whether to proceed with the protocol
or discard. In a semi-quantum scheme, a classical users cannot produce
decoy qubits, so to securely transmit $n$-bit of classical information,
he/she must receive $2n$ qubits from the quantum user, who would
require to send these 2$n$-qubits to be used by user along with another
$2$$n$ qubits, which are decoy qubits for the quantum user- classical
user transmission. Thus, quantum user need to prepare and send $4n$
qubits to a classical user for sending $n$ bits of classical communication. 

\begin{center}
\begin{table}[H]
\begin{centering}
\begin{tabular}{|c|c|c|c|c|c|c|>{\centering}p{2cm}|}
\hline 
Protocol & Task & $c$ (bits) & $q_{c}$ (qubits) & $d$ (qubits) & $q=q_{c}+d$ (qubits) & $b$ (bits) & Qubit efficiency $\eta=\frac{c}{q+b}$\tabularnewline
\hline 
Protocol 1 & SQKA & $n$ & $2n$ & $3n$ & $5n$ & $5n$ & $\eta=10\%$\tabularnewline
\hline 
Protocol 2 & CDSSQC & $n$ & $4n$ & $13n$ & $17n$ & $8n$ & $\eta=4\%$\tabularnewline
\hline 
Protocol 3 & CDSSQC & $n$ & $3n$ & $10n$ & $13n$ & $8n$ & $\eta=4.72\%$\tabularnewline
\hline 
Protocol 4 & SQD & $2n$ & $2n$ & $3n$ & $5n$ & $5n$ & $\eta=20\%$\tabularnewline
\hline 
\end{tabular}
\par\end{centering}

\textcolor{blue}{\caption{\label{tab:The-qubit-efficiency}The qubit efficiency of proposed
semi-quantum protocols.}
}
\end{table}

\par\end{center}

The details of the number of qubits used for both sending the message
($q_{c}$) and checking an eavesdropping attempt ($d$) in all the
protocols proposed here are explicitly mentioned in Table \ref{tab:The-qubit-efficiency}.
In the last column of the table the computed qubit efficiencies are
 listed. From Table \ref{tab:The-qubit-efficiency}, one can easily
observe that the qubit efficiency of three party schemes (Protocol
2 and 3) is less than that of two party schemes (Protocol 1 and 4).\textcolor{blue}{{}
}This can be attributed to the nature of three party schemes, where
one party is supervising the one-way semi-quantum communication among
two remaining parties, which increases the resource requirements to
ensure the control power. In the controlled semi-quantum schemes,
the qubit efficiency computed for Protocol 3 is comparatively more
than Protocol 2 as the controller had chosen to prepare a bipartite
entangled state instead of tripartite entangled state used in Protocol
2. This fact is consistent with some of our recent observations \cite{Voting1}.
Among the two party protocols, Protocol 4 has a higher qubit efficiency
than that of Protocol 1 as the quantum communication involved in this
case is two-way.

Further, one may compare the calculated qubit efficiency with that
of the set of protocols designed for the same task with all the parties
possessing quantum resources. Such a comparison reveals that the requirement
of unconditional security leads to decrease in the qubit efficiency
for the schemes that are performed with one or more classical user(s)
(for example, the qubit efficiency of a QKA scheme was 14.29\% which
is greater than 10\% qubit efficiency obtained here for the SQKA protocol).

\section{Conclusion \label{sec:Conclusion}}

A set of schemes are proposed for various quantum communication tasks
involving one or more user(s) possessing restricted quantum resources.
To be specific, a protocol for key agreement between a classical and
a quantum party is proposed in which both parties equally contribute
in determining the final key and no one can control that key. To the
best of our knowledge this is the first attempt to design a key agreement
between classical and quantum parties. We have also proposed two novel
schemes for controlled communication from a classical sender and a
quantum receiver. It is important to note here that the proposed schemes
are not only the first schemes of their kind, i.e., semi-quantum in
nature, are also shown to be useful in designing a semi-quantum e-commerce
schemes that can provide unconditional security to a classical buyer.\textcolor{blue}{{}
}The presented semi-quantum e-commerce schemes are relevant as the
buyer is supposed to possess minimum quantum resources, while the
online store (the controller) and the merchant may be equipped with
quantum resources. This kind of a semi-quantum scheme can be used
as a solution to a real life problem as in daily life, end users are
expected to be classical. The present work is also the first attempt
for designing semi-quantum schemes having direct real life application.
Further, the first and an unconditionally secure scheme for dialogue
among a classical and a quantum users is proposed here. Later it has
also been shown that the proposed SQD scheme can be modified to obtain
a solution for private comparison or socialist millionaire problem.
The security of the proposed schemes against various possible attack
strategies are also established.

The possibility of realizing semi-quantum schemes for these tasks
establishes that the applicability of the idea of semi-quantum schemes
is not restricted to key distribution and direct communication between
a classical and quantum or two classical (only with the help of a
third quantum party) parties. Our present work not only shows that
almost all the secure communication tasks that can be performed by
two quantum parties can also be performed in a semi-quantum manner
at the cost of increased requirement of the quantum resources to be
used by the quantum party. To establish this point, the qubit efficiencies
of the proposed schemes are computed, which are evidently lower than
the efficiency of the similar schemes with all the parties possessing
quantum resources. 

With the recent development of experimental facilities and a set of
recent experimental realizations of some quantum cryptographic schemes,
we hope that the present results will be experimentally realized in
the near future and these schemes or their variants will be used in
the devices to be designed for the daily life applications.

\textbf{\noindent Acknowledgment:} CS thanks Japan Society for the
Promotion of Science (JSPS), Grant-in-Aid for JSPS Fellows no. 15F15015.
KT and AP thank Defense Research \& Development Organization (DRDO),
India for the support provided through the project number ERIP/ER/1403163/M/01/1603.

\end{document}